
\font\subtit=cmr12
\font\name=cmr8

\input harvmac
\def\UWrLMU#1#2#3#4
{\TITLE{U.T.F. #1}
{IFT UWr #2/\number\yearltd}{#3}{#4}}
\def\TITLE#1#2#3#4{\nopagenumbers\abstractfont\hsize=\hstitle\rightline{#1}
\vskip 1pt\rightline{#2}
\vskip 1in
\centerline{\subtit #3}
\vskip 1pt
\centerline{\subtit #4}\abstractfont\vskip .5in\pageno=0}
\UWrLMU{333}{879}
{OPERATOR FORMALISM ON GENERAL ALGEBRAIC CURVES
}{}
\centerline{F. F{\name ERRARI}$^{a}$,
J. S{\name OBCZYK}$^b$}
\smallskip
$^a${\it Dipartimento di Fisica, Universit\`a di Trento, 38050 Povo (TN),
Italy and INFN, Gruppo Collegato di Trento, E-mail:
francof@galileo.science.unitn.it
}\smallskip
$^b${\it Institute for Theoretical Physics, Wroc\l aw University, pl.
Maxa Borna 9, 50205 Wroc\l aw, Poland,
E-mail: jsobczyk@proton.ift.uni.wroc.pl}\smallskip
\vskip 2cm
\centerline{ABSTRACT}
{\narrower
The usual Laurent expansion of the analytic
tensors on the complex plane is generalized to any closed and
orientable Riemann surface represented as an affine
algebraic curve.
As an application, the operator
formalism for the $b-c$ systems
is developed. The physical states are
expressed by means of creation and annihilation operators as in the complex
plane and the correlation functions are evaluated starting from simple normal
ordering rules. The Hilbert
space of the theory exhibits an interesting internal structure,
being splitted into $n$ ($n$ is the number of branches of the curve)
independent Hilbert spaces.
Exploiting the operator formalism a large collection
of explicit formulas of string theory is derived.
}
\Date{September 1994}

\newsec { INTRODUCTION}
\vskip 1cm
In this paper the Laurent expansion of analytic tensors
defined on the complex plane
is generalized to any algebraic curve associated to a Weierstrass
polynomial of the kind $F(z,y)=0$, with $z\in${\bf CP}$_1$.
The elements of the basis in which the tensors are expanded are explicitly
represented in terms of the multivalued
function $y(z)$ and of the Weierstrass polynomial
itself.
As an application, the fermionic
$b-c$ systems \ref\fms{D. Friedan, E. Martinec and
S. Shenker, {\it Nucl. Phys.} {\bf B271} (1986), 93.}
with integer spin $\lambda\ge 2$ are
quantized
expanding the fields in generalized Laurent series and
requiring that the coefficients of the series
are annihilation and creation
operators. This establishes an operator formalism for the $b-c$ systems
on general algebraic curves and eventually,
due to the conformal equivalence between
Riemann surfaces and algebraic curves
\ref\cenr{F. Enriques and O. Chisini, {\it Lezioni sulla Teoria
Geometrica delle Equazioni e delle Funzioni Algebriche}, Zanichelli, Bologna
(in italian).}, on any Riemann surface.
It is thus remarkable that we are able to derive here all the correlation
functions of the $b-c$ systems starting
from simple commutation relations between operators
on the complex plane.
Another surprising feature of the
$b-c$ systems, which we will show in this work, is that the physical space
of the $b-c$ theory splits into a finite number of independent
Hilbert spaces. For each
Hilbert space it is possible to define a separate vacuum and the creation and
annihilation operators acting on different Hilbert spaces do commute.
In this way we are able to answer with explicit examples the longstanding
question (see e.g. \ref\ceh{A. L. Carey, M. G. Eastwood and K. C. Hannabus,
{\it Comm. Math. Phys.} {\bf 130} (1990), 217.} and references
therein) about the structure of the Hilbert
space in free conformal field theories defined on Riemann
surfaces.
The splitting of the Hilbert space is also
crucial for
proving that there are
no spurious singularities
in the physical amplitudes of the $b-c$ systems.
This was the main obstacle in extending
the approach of refs.
\ref\ffstr{F. Ferrari,
{\it Int. Jour. Mod. Phys.} {\bf A5} (1990), 2799.} and
\ref\janstr{J. Sobczyk and W. Urbanik,
{\it Lett. Math. Phys.} {\bf 21} (1991), 1;
J. Sobczyk, {\it Mod. Phys. Lett.} {\bf A6}
(1991), 1103; {\it ibid.} {\bf A8} (1993), 1153.}
to general algebraic curves
of genus greater than four like
those treated in this work.\smallskip
One of the most important
results obtained from our operator formalism is that we
provide
explicit formulas for the ghost amplitudes in bosonic string theory.
Starting from the original paper of \fms, these amplitudes
have been computed and generalized on any Riemann surface using
different techniques.
A partial list of these works, which in part
exploit and extend the
formulas of \ref\fay{J. D. Fay, Theta Functions on Riemann
Surfaces, Lecture Notes in Mathematical Physics no. 352, Springer
Verlag, 1973.}, is contained in refs.
\ref\vv{E. Verlinde and H. Verlinde, {\it Nucl. Phys.}
{\bf B288} (1987), 357; M. Bonini and R. Iengo, {\it Int. Jour. Mod. Phys.}
{\bf A3} (1988), 841.}, \ref\eo{T. Eguchi and H. Ooguri, {\it Phys. Lett.}
{\bf B187} (1987), 127.}, \ref\abmn{L. Alvarez-Gaum\'e, J.-B. Bost, G.
Moore, P. Nelson and C. Vafa, {\it Comm. Math. Phys.} {\bf 112}
(1987), 503.}, \ref\agr{L. Alvarez-Gaum\'e, C. Gomez and C. Reina, New
Methods in String Theory, in: Superstrings '87, L. Alvarez-Gaum\'e
(ed.), Singapore, World Scientific 1988; N. Kawamoto, Y. Namikawa, A.
Tsuchiya and Y. Yamada, {\it Comm. Math. Phys.} {\bf 116} (1988),
247.}, \ref\blmr{L. Bonora, A. Lugo, M. Matone and J. Russo,
{\it Comm. Math. Phys.} {\bf 123} (1989), 329.}, \ref\raina{A. K. Raina,
{\it Comm. Math. Phys.} {\bf 122} (1989), 625; {\it ibid.} {\bf 140}
(1991), 373; {\it Lett. Math. Phys.} {\bf 19} (1990), 1;
{\it Expositiones Mathematicae} {\bf 8} (1990), 227; {\it Helvetica
Physica Acta} {\bf 63} (1990), 694; A review of an algebraic geometry
approach to a model quantum field theory on a curve, based on the
invited talks in the
conferences: {\it Topology of Moduli Space of Curves}, Kyoto, September 1993,
{\it Vector Bundles on Curves - New Directions}, Bombay and Madras, December
1993, {\it International Colloqium on Modern Quantum Field Theory II}
Bombay, January 1994},
\ref\semi{A. M.
Semikhatov, {\it Phys. Lett.} {\bf B212} (1988), 357; P. di Vecchia,
{\it Phys. Lett.} {\bf B248} (1990), 329; O. Lechtenfeld, {\it Phys.
Lett} {\bf B232} (1989) 193; U. Carow-Watamura, Z. F. Ezawa, K.
Harada, A. Tezuka and S. Watamura, {\it Phys. Lett.} {\it B227}
(1989), 73.}. Our approach is different and it
is based on the representation of Riemann surfaces as $n-$sheeted coverings
of the complex plane. The advantage of this choice is that
the correlation functions can be written in a
simple way in terms of the variables $z$ and $y(z)$ which parametrize
the algebraic curve. Technically speaking, within our formalism it is
possible to achieve what in refs.
\vv\ is called the fermionic construction of the amplitudes, explicitly
taking care of the zero modes also when $\lambda\ge 2$.
It is important to stress at this point that the
representation of
Riemann surfaces as algebraic curves provides interesting
perturbative results in string theory, see for example refs.
\ref\hyper{E. Gava, R. Iengo and C. J. Zhu, {\it Nucl. Phys.} {\bf323} (1989),
585; E. Gava, R. Jengo and G. Sotkov, {\it Phys. Lett. } {\bf 207B}
(1988), 283; R. Jengo and C. J. Zhu, {\it Phys. Lett.} {\bf 212B} (1988), 313;
D. Lebedev and A. Morozov, {\it Nucl. Phys.} {\bf B302} (1986), 163;
A. Yu. Morozov and A. Perelomov, {\it Phys. Lett} {\bf 197B} (1987), 115;
D. Montano, {\it Nucl. Phys.} {\bf B297} (1988), 125.} in the
case of hyperelliptic curves.
Nevertheless, we would also like to emphasize
the strong connections established by our work between
conformal field theories on the complex plane and on Riemann
surfaces. The investigations in this direction started
in refs. \ref\knirev{V. G. Knizhnik, {\it Comm. Math. Phys.}
{\bf 112} (1987), 587; {\it Sov. Phys. Usp.} {\bf 32}(11)
(1989) 945.}, \ref\brzn{
M. A. Bershadsky and A. O. Radul, {\it Int. Jour. Math.
Phys.} {\bf A2} (1987) 165.} in
the case of algebraic curves with $Z_n$ symmetry. In refs.
\ref\plb{F. Ferrari, {\it Phys. Lett.} {\bf B277} (1992), 423.},
\ref\cmp{F. Ferrari, {\it Comm. Math. Phys.} {\bf 156} (1993), 179.}
and \ref\ijmp{F. Ferrari, {\it Int. Jour. Mod. Phys.} {\bf A9} (1994),
313.} these results have been extended to the algebraic curves
with nonabelian group of symmetry $D_n$, showing that an interesting braid
group statistics arises in the amplitudes of the $b-c$ systems.
\smallskip
Despite of the fact that here
we are mainly focused in the construction of the Hilbert
space of the $b-c$ systems, there are many other possible
applications of our operator formalism, for instance in integrable
models and conformal field theories.
In particular, the already mentioned splitting of the
Hilbert space on which the $b-c$ fields act, gives also important insights
in the
Hilbert space of more general conformal field theories.
As a matter of fact, bosonizing the $b-c$ systems, one obtains a theory
of free scalar fields, which is at the  heart of the Coulomb gas
representation of many conformal field theories.
The bosonization program has been completed until now
only in the case of the $Z_n$
symmetric curves
\ref\fsu{F. Ferrari, J. Sobczyk and
W. Urbanik, {\it Operator Formalism on the $Z_n$ Symmetric Algebraic Curves},
Preprint LMU-TPW 93-20, ITP UWr 856/93.},
but can be extended also to the $D_n$
symmetric curves of ref.
\cmp.
Moreover, exploiting the techniques introduced in this work,
we hope that it will be possible to bosonize
the $b-c$ systems also on more general curves.
Concerning integrable models, the
basis given here for the $b-c$ fields generalizes
an analogous one
introduced in \ref\pp{S. Pakuliak and A. Perelomov, {\it Relations
Between Hyperelliptic Integrals}, Preprint DFTUZ/94/07, hep-th/9505018.}
in order to explain why the solutions of
the level zero ${sl(2)}$ Knizhnik-Zamolodchikov
equations
are given in terms of hyperelliptic integrals.
With our basis it seems possible to extend the results
of ref. \pp\ also
to the level zero ${sl(n)}$ Knizhnik-Zamolodchikov equations,
which are satisfied by the form factors of the $su(n)-$invariant Thirring
model and whose solutions are related to the periods of differentials
on algebraic curves with $Z_n$ group
of symmetry \ref\fas{F. A. Smirnov, {\it Comm. Math. Phys.}
{\bf 155} (1993),459.}. The explicit
formulas given in this paper can also shed some light
on the conjectured connections between
general algebraic curves and integrable models \fas, \ref\matve{V. Matveev,
Deformations of Algebraic Curves and Integrable Nonlinear Evolution Equations,
Preprint MPI-PTH/14-91.}, \ref\matvee{ The only explicit example known by us
of the connections between curves with nonabelian monodromy group and
integrable models is given in:
D. A. Korotkin and V. B. Matveev, {\it Leningrad Math. Jour.} {\bf 1} (1990),
379.}.
Finally, there is a fascinating and
not yet proved
suggestion
by Bershadsky-Knizhnik-Radul
\knirev, \brzn, that the
$b-c$ systems
defined on algebraic curves
might be equivalent to conformal field theories
defined on the complex plane.
Using the bosonized version of our formalism, this equivalence can
possibly be
verified as it has been shown in the particular case of the
$Z_n$ curves \fsu.
\smallskip
This paper is organized as follows.
In section 2 the usual Laurent expansions are extended
to general algebraic curves
of genus $g$.
In order to point out that there is a superposition of independent
modes in the physical amplitudes of the $b-c$ systems,
which leads to the splitting of the Hilbert space, two different Laurent
expansions are introduced for the $b$ and $c$ fields separately.
The completeness and equivalence of both bases in the space of the
analytic tensors
defined on $\Sigma_g$ is shown in Proposition 1.
Furthermore, the elements of the two bases
are explicitly
derived in terms of the coordinates $z$ and $y(z)$ and their
divisors are computed as well.
For technical simplicity, for example
in the computation of the genus and of the relevant divisors,
we consider nondegenerate algebraic curves, but the outcomes will turn out to
be completely general.
In section 3 we introduce the operator
formalism for the $b-c$ systems with $\lambda>1$.
The case of $\lambda=1$ is complicated by the presence of the zero mode
in the field $c$ and will be treated in a separate paper.
We verify the splitting of the Hilbert space in a finite number of
independent sectors. The normal ordering
between the fields is a simple generalization of the usual normal ordering
on the complex plane. All the correlation functions of the $b-c$ systems
on an algebraic curve are then obtained as vacuum expectation values
of multivalued fields.
In Section 4 we check the operator formalism computing the $n-$point
functions of the $b-c$ systems on a quartic of genus three and on the
$Z_n$ symmetric curves. These curves represent,
together with other few examples, the only cases in which
the results are already known by exploiting other methods
\ffstr, \janstr, \knirev, \brzn.
Finally, in section 5 we present the conclusions and discuss
the possible future developments.
\vskip 1cm
\newsec{GENERALIZED LAURENT EXPANSIONS ON ALGEBRAIC CURVES}
\vskip 1cm
We consider here the (class of) algebraic curves $\Sigma_g$
associated to the
vanishing of the following Weierstrass polynomial:

\eqn\curve{
F(z,y)=P_n(z)y^n+P_{n-1}(z)y^{n-1}+ ... +P_1(z)y+P_0(z)=0
}
where the $P_j(z)=\sum\limits_{m=0}^{n-j}\alpha_{j,m}z^m$
are polynomials in $z$ of degree at most $n-j$,
$j=0,\ldots,n$ and the $\alpha_{j,m}$ are
complex parameters. $y$ can be viewed as a multivalued function
on the complex sphere {\bf CP}$_1$ or as a meromorphic function
on the  Riemann surface defined by eq. \curve.
The function $y(z)$ has poles only at $z=\infty$.
By suitably setting to zero some of the parameters $\alpha_{j,m}$
it is possible to obtain every other kind of Weiestrass polynomials.
However, in order to fix the ideas,
though it will be not strictly necessary in the following,
we suppose that the algebraic curves treated here are nondegenerate.
We recall that, in order to define nondegenerated curves, one has to introduce
an homogeneous polynomial $\tilde F(z,y,w)$ of degree $n$
in the three variables $z$, $y$ and $w$
defined in such a way that $\tilde F(z,y,1)=F(z,y)$.
In our case it is sufficient to change the polynomials
$P_j(z)$'s into $\tilde P_j(z,w)=\sum_{m=0}^{n-j}\alpha_{j,m}z^m w^{n-j-m}$.
The nondegeneracy condition amounts to the requirement that
the equations:
\eqn\degeneracy{\tilde F(z,y,w)=\tilde F_y(z,y,w)=\tilde F_z(z,y,w)=\tilde
F_w(z,y,w)=0}
where $\tilde F_y(z,y,w)\equiv\partial_w \tilde F(z,y,w)$,
$\tilde F_z(z,y,w)\equiv\partial_z \tilde F(z,y,w)$
and
$\tilde F_w(z,y,w)\equiv\partial_w \tilde F(z,y,w)$
are never simultaneously satisfied.\medskip
Since eq. \degeneracy\
is an overdetermined system of algebraic equations, it can be
satisfied only for very particular choices of the parameters $\alpha_{j,m}$.
Therefore,
the nondegenerate solutions of eq. \curve\
still describe an huge class of algebraic curves.
Remarkably, the genus of the equivalent Riemann surface can
be directly computed from the form of the Weiestrass polynomial
\curve. This is the Baker's method, explained in ref.
\ref\forsyth{A. R. Forsyth, Theory of Functions of a Complex
Variable, Vols. I and II, Dover Publications Inc., New York, 1975.}
Vol. I, p. 404.
It turns out that the genus of $\Sigma_g$ is
$g={(n-1)(n-2)\over 2}$.\smallskip
On the algebraic curves \curve\ we consider a theory of
free fermionic
$b - c$ systems with integer spin $\lambda\ge 2$ defined by the following
action:
\eqn\action{S_{\rm bc}=\int_{\Sigma_g}d^2\xi\left(b\bar\partial c+{\rm
c.c.}\right)}
where $\xi$ and $\bar \xi$ are complex coordinates on $\Sigma_g$.
An introduction of the effective operatorial formalism should follow from
the suitable identification of the classical degrees of freedom of the fields
$b$ and $c$. For that purpose let us analyze the solutions of the classical
equations of motion descending from eq. \action:

\eqn\classeq{\bar\partial b=\bar\partial c=0}

We shall expand them in the following basis:

\eqn\bdz{b(z)dz^\lambda=\sum\limits_{k=0}^{n-1}\sum\limits_{i=-\infty}
^\infty b_{k,i}z^{-i-\lambda}f_k(z)dz^\lambda}

\eqn\cdz{c(z)dz^{1-\lambda}=
\sum\limits_{k=0}^{n-1}\sum\limits_{i=-\infty}^\infty
 c_{k,i}z^{-i+\lambda-1}\phi_k(z)dz^{1-\lambda}}

with $f_k$ and $\phi_l$ chosen as follows
($k,l=0, ..., n-1$):

\eqn\fkn{
f_k(z) = {y^{n-1-k}(z)dz^{\lambda}\over (F_y(z,y(z)))^{\lambda} }
}
$$
\phi_l(w) dw^{1-\lambda}
= { dw^{1-\lambda}\over(F_y(w,y(w)))^{1-\lambda}}\times
$$
\eqn\phikn{
\left( y^l(w)+y^{l-1}(w)P_{n-1}(w)+y^{l-2}(w)P_{n-2}(w)+...+P_{n-l}(w)
\right)
}
We notice that we have introduced two different expansions for
the fields $b$ and $c$. This is only for the sake
of convenience in the formulation
of the operatorial formalism. In fact, after
the replacement $\lambda\rightarrow 1-\lambda$, the basis
\eqn\basisone{B_{i,k}(z,y(z))
dz^\lambda=z^i f_k(z)dz^\lambda\qquad\qquad\qquad\cases{i=0,1,\ldots\cr
k=0,\ldots,n-1\cr}}
used in eq. \bdz\ for the $\lambda-$differentials turns out to be equivalent
to the basis
\eqn\basistwo{C_{i,k}(w,y(w))dw^{1-\lambda}=z^i \phi_k(w)
dw^{1-\lambda}\qquad\qquad\qquad\cases{i=0,1,\ldots\cr
k=0,\ldots,n-1\cr}}
used in the case of the $1-\lambda$ differentials.
In other words, the elements $B_{i,k}(z,y(z))
dz^\mu$ are linear combinations of the $C_{i,k}(z,y(z))
dz^{1-\lambda}$ and viceversa if $\mu=1-\lambda$ as it is easy to prove.
The reasons of taking asymmetric expansions for the fields $b$ and $c$
will be clear when the correlation functions
are computed, showing that they
are multilinear superpositions of the elements
$B_{i,k}(z,y(z))
dz^\lambda$ and $C_{i,k}(z,y(z))
dz^{1-\lambda}$
For the moment, we just state the following proposition, which will be proved
in Appendix A:
\smallskip\noindent
\item{}{\bf Proposition 1} {\it
Every $\lambda-$differential $\omega dz^\lambda$, $\lambda\in${\bf Z},
on an arbitrary algebraic
curve $\Sigma_g$ can be expanded in terms
of the basis whose elements are given by eq. \basisone\ as follows:
\eqn\proppone{\omega(z)dz^\lambda=\sum\limits_{k=0}^{n-1}g_k(z)f_k(z,y(z))
dz^\lambda}
where $f_k(z,y(z))$ has been defined in eq. \fkn\ and the $g_k(z)$'s are
singlevalued, rational functions of $z$. An analogous statement
is also true for the basis \basistwo.
}\smallskip\noindent
Hence, eqs. \bdz\ and \cdz\ can be regarded as the generalization
on algebraic curves of the standard
Laurent expansion on the complex plane. This is a great advantage with respect
to the usual Poiseaux expansions, defined in the
neighborhood of a point and, consequently, only locally valid.
If the $P_j(z)$ are set to zero for
$j=1,\ldots,n-1$, the limiting case of $Z_n$ symmetric
algebraic curves is obtained. It is then possible to check that the expansions
\bdz\ and \cdz\ become equal to
those obtained in ref. \fsu.\smallskip
In order to ascertain that the physical amplitudes do not
contain spurious poles and to compute the zero modes, it is necessary to
determine the analytical properties of the elements
of the bases \basisone\ and \basistwo. The best strategy is to
derive first the divisors of their building blocks, i.e. of $dz$, $y$ and
$F_y(z,y)$.
These divisors can be easily evaluated on any algebraic curve using
the methods of ref. \ffstr, but unfortunately there is no universal procedure
which applies automatically to any kind of Weierstrass polynomials,
in particular to those corresponding to degenerate curves.
The branch points of the curve
$\Sigma_g$ are determined by the conditions:
\eqn\brpo{F(z,y)=F_y(z,y)=0}
Eliminating
$y$ from the two equations written above one finds an
equation in the
variable $z$ of the kind $r(z)=0$.
$r(z)$ is the resultant of the system \brpo\ and it is a polynomial
expressed in terms of the $P_j(z)$, $j=0,\ldots,n$, appearing in eq.
\curve.
This can be seen using for instance
the dialitic method of Sylvester
(see ref. \cenr, Vol. II, p. 79).
Performing the calculations explicitly
it turns out that, in general, the
degree of $r(z)$  is equal to $n(n-1)$.
In this case there are no branch points at infinity as we shall suppose
throughout this paper for the sake of simplicity.
This implies that there are $n_{bp}$ finite
branch points $a_1,\ldots,a_{n_{bp}}$
of multiplicity $\nu_i$, where $\nu_i$
describes the number of branches of $y$ connected
at the branch point $a_i$. The multiplicities and the number of branch
points  satisfy the following Riemann$-$Hurwitz equation:
\eqn\rh{2g-2=-2n+\sum\limits_{i=1}^{n_{bp}}(\nu_i-1)}
The algebraic curves in which all the branch points
have multiplicity two are said to be in the normal form and are
well known
in the mathematical literature \forsyth, \cenr.
For example, the corresponding Riemann surfaces can be
explicitly constructed in terms of branch points and branch lines
and even a basis of independent homology
cycles is known (L\"uroth Theorem).
It is clear that surfaces in the normal
form are particular cases of our general scheme.
\smallskip
At this point, denoting the branch points by $a_i$, $i=1,\ldots,n_{bp}$
and following the methods explained in ref. \ffstr,
we obtain the desired
divisors:
\eqn\divdz{
[dz]=\sum^{n_{bp}}_{p=1}(\nu_p-1) a_p\ -\ 2\sum^{n-1}_{j=0}\infty_j
}
\eqn\divy{
[y]=\sum^n_{r=1}q_r\ -\ \sum^{n-1}_{j=0}\infty_j
}
\eqn\divdf{
[F_y]=\sum^{n_{bp}}_{p=1}(\nu_{p}-1)a_p\ -\ (n-1)\sum^{n-1}_{j=0}\infty_j
}
In eq. \divy\ the $q_j$ denote the zeros of $y$ which, when projected from
the Riemann surface on the $z$ complex plane, coincide with the
zeros of $P_0(z)$.
Moreover, $\infty_j$
describes the projection of the point at infinity
on the $j$-th sheet of the Riemann surface. Finally, in our conventions
positive and negative integers denote the order of the zeros and of
the poles respectively.
Starting from eqs. \divdz-\divdf\ it is possible to find the
divisors of the elements of the basis \basisone:
\eqn\divf{
[z^if_k]=(n-1-k)\sum^n_{s=1}q_s + i\sum^{n-1}_{l=0} 0_l
+(k+\lambda (n-3) - (n-1)-i)\sum^{n-1}_{l=0}\infty_l
}
where, using the notation exploited for the point at infinity,
$0_l$ denotes the projection of the point $z=0$ on the $j-th$ sheet.
An analogous formula can be derived for the
$C_{i,k}(w,y(w))dz^{1-\lambda}$.
\smallskip
To conclude this section
we explicitly
derive the form of the zero modes associated to the
equations of motion \classeq.
We try the following ansatz:
\eqn\zm{
\Omega_{k,i}dz^\lambda=f_k(z)z^{-i-\lambda}
}
After the substitution
$i\rightarrow -i-\lambda$ in eq.
\divf, we see that $\Omega_{k,i}dz^\lambda$
has no singularities whenever
\eqn\zma{
i\leq -\lambda
}
and
\eqn\zmb{
k+\lambda (n-2) - (n-1) +i \geq 0
}
In order to find all the possible zero modes we proceed by fixing the
value of $k$ in eq. \zmb\ and then computing the possible values
of $i$ compatible also with eq. \zma.
Skipping the simple cases in which $n=2,3$ corresponding to genus zero
and one Riemann surfaces, we obtain the following results:\medskip
\item{1)} $n>4$, $\lambda>1$ or $n=4$, $\lambda>2$.
Then the $N_b=(2\lambda-1)(g-1)$ independent zero modes
are of the form given by eq. \zm\ and:
\eqn\caseone{\cases{k=0,\ldots,n-1\cr
\lambda(2-n)+n-1-k\le i\le -\lambda\cr}}
Moreover the number of zero modes proportional to $f_k$ is given by:
\eqn\nbkone{N_{b_k}=\lambda(n-3)+k-n+2}
\medskip
\item{2)} $n=4$, $\lambda=2$. In this case the genus of the curve is three
and the six
independent zero modes occur when
\eqn\casetwo{\cases{k=1,2,3\cr
-1-k\le i\le -2}}
and
\eqn\nbktwo{N_{b_k}=k\qquad\qquad\qquad k=1,2,3}
\medskip
\item{3)} $n>3$, $\lambda=1$. Here the conditions determining all the
$g$ independent zero modes
of the form \zm\ are:
\eqn\casethree{\cases{k=2,3,\ldots,n-1\cr
-k+1\le i\le -1\cr}}
while
\eqn\nbkthree{N_{b_k}=k-1\qquad\qquad\qquad k=2,\ldots,n-1}
\medskip
\item{4)} Finally, if $\lambda=0$, we have only one zero mode provided
by the constant function in the
$c$ fields. Using the basis \basistwo\ this zero mode is obtained
for $k=i=0$ .
\medskip\noindent
In the first two cases written above it is
easy to check that total number of zero modes is
\eqn\zmc{
N_b=\sum_{k=0}^{n-1}N_{b_k}={(2\lambda -1)(n-3)n\over 2}
}
which in fact expresses the well known Riemann-Roch theorem
for a surface of genus $g$.
\vskip 1cm
\newsec{THE OPERATOR FORMALISM}
\vskip 1cm
In the previous section we have shown how to expand
the classical $b-c$ fields on an algebraic curve in generalized Laurent
series.  Now we quantize the fields transforming the coefficients
$b_{i,k}$ and $c_{i,k}$ of eqs. \bdz\ and \cdz\ into creation and
annihilation operators.
In the following, it will be convenient to exploit the basis \basisone\
in order to expand the $b$ fields
for $\lambda>1$ and the basis \basistwo\ in order to expand
the $c$ fields. This is just a matter of convenience since the bases
\basisone\ and \basistwo\ have no particular distinguishing
properties. In principle, one
could also expand the $b-c$ fields in terms of another
basis, but
this would require a certain amount of complications since the zero
modes and all the other tensors entering the theory are most
naturally expressed
as linear combinations of the elements
\basisone\ and \basistwo.
At this point we divide the degrees of freedom of
the $b-c$ systems into $n$ sectors,
numbered by the index $k=0,\ldots,n-1$ and
characterized
by the tensors $f_k$ for the $b$ fields and by the tensors
$\phi_k$ for the $c$ fields. This seems a natural choice, since
all the zero modes are expressed in a simple way
in terms of $f_k$. In fact, we have seen in the previous
section that it is even possible to define
the numbers $N_{b_k}$ of the
zero modes corresponding to $f_k(z)$.
Further, we
assume that in our operator formalism
the modes $z^jf_k(z)$
labelled by different indices $k$ do not interact.
This hypothesis, to be proven a posteriori,
implies that the space on which the $b-c$ fields defined on a Riemann
surface act can be decomposed
into a set of $n$ independent Hilbert spaces if the Riemann surface is
represented as an $n-$sheeted branch covering over {\bf CP}$_1$.
Accordingly, we
quantize the theory \action\ postulating the following basic anticommutation
relations for the coefficients $b_{k,i}$ and $c_{k,i}$ appearing in
eqs. \bdz\ and \cdz\ respectively:
\eqn\commrel{\{b_{k,i},c_{k',i'}\}=\delta_{kk'}\delta_{i+i',0}.}
For each value of $k=0,...,n-1$  these creation and annihilation operators act
on the vacuum $|0\rangle_k$, $k=0,\ldots,n-1$,
which represents the usual vacuum of the $b-c$ systems
on the complex plane.
The ``total vacuum" of the $b-c$ systems on $\Sigma_g$ is given by

\eqn\totalvacuum{|0\rangle=\otimes_{k=0}^{n-1}|0\rangle_k.}

We demand that

\eqn\ban{b^-_{k,i}|0\rangle_k\equiv b_{k,i}|0\rangle_k=0
\qquad\qquad\qquad\left\{\eqalign{k=&0,\ldots,n-1\cr i
\ge& 1-\lambda\cr}\right.}

\eqn\can{c^-_{k,i}|0\rangle_k\equiv c_{k,i}|0\rangle_k=0
\qquad\qquad\qquad\left\{\eqalign{k=&0,\ldots,n-1\cr i\ge&
\lambda\cr}\right.}

Moreover. we introduce the "out" vacua ${}_k\langle 0|$ requiring that

\eqn\bcrea{{}_k\langle0| b^+_{k,i}\equiv {}_k\langle 0|b_{k,i}=0
\qquad\qquad\qquad\left\{\eqalign{k=&0,\ldots,n-1\cr i
\le& -\lambda-N_{b_k}\cr}\right.}
\eqn\ccrea{{}_k\langle 0|c^+_{k,i}\equiv {}_k\langle 0|c_{k,i}=0
\qquad\qquad\qquad\left\{\eqalign{k=&0,\ldots,n-1\cr i\le&
\lambda-1\cr}\right.}
{}From the above equations we see
that some operators $b_{k,j}$ correspond to zero modes and the remaining
ones are organized in two sets of creation and annihilation ones.
The annihilation operators annihilate states with negative energy
as it is possible to verify from eq. \bdz.
The same applies to the $c_{k,j}$
with the only difference that there are no zero modes for them.\smallskip
Finally, let us
introduce the following useful notations

\eqn\bkdz{b_k(z)dz^\lambda=f_k(z)
\sum\limits_{i=-\infty}^{\infty}b_{k,i}
z^{-i-\lambda}dz^\lambda}
\eqn\ckdz{c_k(z)dz^{1-\lambda}=\phi_k(z)
\sum\limits_{i=-\infty}^\infty
c_{k,i}z^{-i+\lambda-1}dz^{1-\lambda}}
\eqn\bdzcdz{b(z)dz^\lambda=\sum_{k=0}^{n-1}b_k(z)dz^\lambda\qquad\qquad\qquad
c(z)dz^{1-\lambda}=\sum_{k=0}^{n-1}c_k(z)dz^{1-\lambda}}
{}From the above considerations, by exploiting the commutation relations
\commrel\ and  the
natural definition of the ``normal ordering" of the $b-c$ systems, we get:

\eqn\bkzckwnorm{b_k(z)c_k(w)=:b_k(z)c_k(w):+{1\over z-w}f_k(z)\phi_k(w)}

The ``time ordering" is implemented by
the requirement that the fields $b(z)$ and $c(w)$
are radially ordered with respect to the variables $z$ and $w$.
Finally,  in order to take into account the zero modes, we impose the following
conditions:
\eqn\vaccond{{}_k\langle 0|0\rangle_k=0\qquad{\rm if}\ N_{b_k}\ne 0;
\qquad\qquad
{}_k\langle 0|\prod\limits_{i=1}^{N_{b_k}}b_{k,i}|0\rangle_k=1.}
Starting from
eqs. \commrel-\vaccond\ it is now possible to compute on any algebraic
curve the correlation functions of the $b-c$ systems as
expectation values
over the vacuum \totalvacuum\
of the fields $b$ and $c$ defined in eqs. \bdzcdz.
However the numbers $N_{b_k}$
of the  zero modes must be individually computed for each different class
of algebraic curves.\smallskip
As a first step, we calculate the following amplitudes:
\eqn\propone{{}_k\langle 0|b_k(z_1)\ldots b_k(z_{N_{b_k}})|0\rangle_k=
{\rm Det}\left|
\Omega_{k,j}(z_i)\right|\qquad\qquad\qquad i,j=1\ldots,N_{b_k}}
\eqn\proptwo{\langle 0|b(z_1)\ldots  b(z_{N_b})|0\rangle={\rm
det}\left|\Omega_I(z_J)\right|}
where the vacuum $|0\rangle$ and the fields $ b(z)dz^\lambda$ have already been
defined in eqs. \totalvacuum\ and \bdzcdz. Moreover $I,J=1,\ldots,
\sum\limits_{k=0}^{n-1}=N_b$, $N_b$ denoting the total number of
zero modes.
Finally the $\Omega_I(z)dz^\lambda$ represent all the possible zero
modes with spin $\lambda$:
$$\Omega_I(z)dz^\lambda\in\left\{\Omega_{k,i}(z)dz^\lambda|1\le i\le
N_{b_k}, 0\le k\le n-1\right\}.$$
The zero modes
$\Omega_{k,j}(z)dz^\lambda$ can be obtained in terms of $z$ and $y$ from eqs.
\zm\ and \fkn:
\eqn\omzero
{
\Omega_{k,j} (z) \equiv { y^{n-1-k}z^{j-1}\over (F_y(z,y(z)))^{\lambda}}
dz^{\lambda}
}
where the range of the indices $k=0, ..., n-1$ and $j=1, ..., N_{b_k}$
is given in eqs. \nbkone\ and \nbktwo.
\smallskip
The proof of eq. \propone\ is straightforward, while eq. \proptwo\ will be
proven in Appendix B. We notice that eq. \proptwo, obtained here in a pure
operatorial way from the
basic commutation relations \commrel\ satisfied by
the creation and annihilation
operators $b_{k,i}$ and $b_{k,i}$, is in complete agreement with the
standard results \vv.

Now we are ready to compute the propagator of the $b-c$ fields, which, in our
formalism, is defined by the following ratio of correlators:

\eqn\propoper{G_\lambda(z,w)={\langle 0| b(z)
c(w)\prod\limits_{I=1}^{N_b}
 b(z_I)
|0\rangle\over
\langle 0|\prod\limits_{I=1}^{N_b}b(z_I)|0\rangle}}
{}From eq. \bkzckwnorm\ the normal ordering of any two
fields
$b$ and $c$ becomes:
\eqn\normord{ b(z) c(w)=
: b(z) c(w):+K_\lambda(z,w)dz^{\lambda}dw^{1-\lambda}.}
where $K_\lambda(z,w)$ denotes the following tensor:
\eqn\ksum
{
(z-w) K_{\lambda}(z,w)=\sum\limits_{j=0}^{n-1} f_j(z)\phi_j(w)}
In order to fix the overall sign in the expression of the propagator
we suppose that the fields contained in the
correlators appearing in eq. \propoper\ are already radially ordered, i.e.
$$|z|>|w|>|z_1|\ldots>|z_{N_b}|.$$
Thus, applying eq. \normord, we obtain:
$${\langle 0| b^{(l)}(z)
c^{(l')}(w)\prod\limits_{I=1}^{N_b} b^{(l_I)}(z_I)
|0\rangle\over
\langle 0|\prod\limits_{I=1}^{N_b} b^{(l_I)}(z_I)|0\rangle}=
K_\lambda^{(ll')}(z,w)dz^\lambda
dw^{1-\lambda}+$$
\eqn\propint{\sum\limits_{J=1}^{N_b}(-1)^JK_\lambda^{(l_Jl')}(z_J,w)
{\langle 0|  b^{(l_1)}(z_1)\ldots
 b^{(l_{J-1})}(z_{J-1}) b^{(l)}(z) b^{(l_{J+1})}
(z_{J+1})
\ldots  b^{(l_{N_b})}(z_{N_b})|0\rangle\over
\langle 0|  b^{(l_1)}(z_1)\ldots  b^{(l_{N_b})}(z_{N_b})|0\rangle}}
where $l$, $l'$ and $l_I,l_J$ denote the branches of fields and tensors
in the variables $z$, $w$, $z_I, z_J$ respectively.
The residual correlation functions in eq. \propint\ contain products of
$N_b$ fields $ b$ and therefore can be easily computed by means of eq.
\proptwo.
The final result is the following propagator:
$${\langle 0| b^{(l)}(z) c^{(l')}(w)\prod\limits_{I=1}^{N_b}
 b^{(l_I)}(z_I)
|0\rangle\over
\langle 0|\prod\limits_{I=1}^{N_b} b^{(l_I)}(z_I)|0\rangle}=$$
\eqn\propfin{{
{\rm det}\left|\matrix{\Omega_1^{(l)}(z)&\ldots&
\Omega_{N_b}^{(l)}(z)&K_\lambda^{(ll')}(z,w)\cr
\Omega_1^{(l_1)}(z_1)&\ldots&
\Omega_{N_b}^{(l_1)}(z_1)&K_\lambda^{(l_1l')}(z_1,w)\cr
\vdots&\ddots&\vdots&\vdots\cr
\Omega_1^{(l_{N_b})}(z_{N_b})&\ldots&
\Omega_{N_b}^{(l_{N_b})}(z_{N_b})&K_\lambda^{(l_{N_b}l')}(z_{N_b},w)
\cr}\right |\over
{\rm det}\left|\Omega_I(z_J)\right|}}
Eq. \propfin\ is the final expression of the propagator of the $b-c$ systems
evaluated on a general algebraic curve.
Of course, we have still to check that
the tensor appearing
in the right hand side of eq. \propfin\ has only a physical
simple pole at the points
$z=w$ and $w=z_J$, $J=1,\ldots,N_b$.
Moreover, since the propagator is a multivalued
tensor, these singularities must occur when the branches
coincide, i. e. when $l=l'$ and $l'=l_J$ respectively.
To verify that the propagator \propint\ has the correct singularities,
we apply the divisors \divy-\divf\ to the tensor $K_\lambda(z,w)$, which is
responsible for the divergences. In this way we restrict ourselves
to the general curves of genus $g=(n-1)(n-2)/2$. However, it
is possible to verify the absence
of spurious poles also for the other curves. Some examples will be reported
in the next section.\smallskip
First of all we treat the case 1) explained in the previous section, i. e.
when eq. \caseone\ is valid.
In this case, apart from the factor $1/(z-w)$,
$K_\lambda(z,w)$ is a linear combination of the
$\lambda-$differentials $f_k(z)$
which are all zero modes.
{}From eq. \divf\ it is in fact easy
to see that
the singularities of $K_\lambda(z,w)$
in the variable $z$ can occur only when $z=w$.
In the variable $w$, instead, in addition to the singularities in
$w=z$ there are spurious poles at infinity given by the $1-\lambda$
differentials $\phi_k(w)$.
Of course these spurious poles must not contribute in the final
expression of the propagator.
To show this, we rewrite $K_\lambda(z,w)$ as follows:
$$K_\lambda(z,w)=\sum_{k=0}^{n-1}u_k(z,w)$$
where
$$u_k(z,w)={1\over z-w}\phi_k(w)f_k(z)$$
We obtain the following
behavior for $u_k(z,w)$ when $w\rightarrow \infty$:
\eqn\ukbehav{u_k(z,w)\sim w^{\lambda(n-3)+k-n+2}+\ldots}
Therefore, there is a danger of spurious poles whenever
$\lambda(n-3)+k-n+2>0$.
Let us call $u_k(z,w)_{\rm div}$ the divergent terms in the
asymptotic expansion at $w=\infty$ of
$u_k(z,w)$.
{}From \ukbehav,
in order to compute $u_k(z,w)_{\rm div}$, we need to expand
the quantities $1/(z-w)$ and $y(w)$ up to the order
$w^{-(1+\lambda(n-3)+k-n+1)}$ and substitute them back in the expression of
$u_k(z,w)$:
\eqn\expone{{1\over z-w}\sim {1\over w}\left(1+{z\over w}+\ldots+\left({z\over
w}\right)^{\lambda(n-3)+k-n+1}\right)+\ldots}
Moreover, since from eq. \divy\ $y(w)$ is not branched in $w=\infty$, we have:
\eqn\exptwo{y(w)\sim\sum^{1}_{s=-(\lambda(n-3)+k-n+2)}
\gamma_sw^s+\ldots}
where the coefficients $\gamma_i$, $i=1,\ldots, -(\lambda(n-3)+k-n+2)$ can
be computed in terms of the parameters $\alpha_{j,m}$ by direct
insertion of the right hand side of the above equation in eq. \curve.
Using the expansions \expone\ and \exptwo\
we obtain that $u_k(z,w)_{\rm div}$ is
of the following form:
$$
u_k(z,w)_{\rm
div}=\left(\beta_{\lambda(n-3)+k-n+2}
w^{\lambda(n-3)+k-n+2}+\beta_{\lambda(n-3)+k-n+1}
w^{\lambda(n-3)+k-n+1}z+\ldots\right.$$
\eqn\ukadiv{\left.+\beta_1w z^{\lambda(n-3)+k-n+1}
\right)f_k(z)}
Eq. \ukadiv\ shows that the divergent part of $u_k(z,w)$ is
proportional
to a $\lambda-$differential in $z$ of the kind
$\Omega_{i,k}dz^\lambda=z^if_k(z)$, $i=0,\ldots,\lambda(n-3)+k-n+1$,
$k=0,\ldots,n-1$.
On the other side, after performing the
substitution $-i-\lambda\rightarrow i$ in eq. \divf,
it is possible to see that the
$\Omega_{i,k}dz^\lambda$ are zero modes if the indices $i$ and $k$ are
taken as above.
As an upshot, looking at eq. \propfin, which is a determinant of a
matrix whose columns contain all the independent zero modes, it turns
out that the contribution given by the terms carrying
the spurious poles at $w=\infty$ in $K_\lambda(z,w)$ is zero:
$${
{\rm det}
\left|
\matrix{
\Omega_1^{(l)}
(z)&\ldots&
\Omega_{N_b}^{(l)}(z)&
u_k^{(ll')}(z,w)_{\rm
div}\cr
\Omega_1^{(l_1)}(z_1)&\ldots&
\Omega_{N_b}^{(l_1)}(z_1)&
u_k^{(l_1l')}(z_1,w)_{\rm div}\cr
\vdots&\ddots&\vdots&\vdots\cr
\Omega_1^{(l_{N_b})}
(z_{N_b})&\ldots&
\Omega_{N_b}^{(l_{N_b})}
(z_{N_b})&
u_k^{(l_{N_b}l')}(z_{N_b},w)_{\rm div}
\cr
}
\right |
\over
{\rm det}
\left|
\Omega_I(z_J)
\right|
}=0$$
This concludes our proof that there are no singularities at $w=\infty$
in the propagator \propfin. However, the proof is valid only if eq. \caseone\
is true. The exception is provided by
the $2-$differentials on an algebraic curve of genus three (see eq. \casetwo).
Taking $n=4$ and $\lambda=2$ in eq. \curve, in fact, we have for $k\ge 1$:
$$u_k(z,w)_{\rm div}=\left(\beta_kw^k+\ldots+\beta_1wz^{k-1}\right)f_k(z)$$
These terms are again proportional to zero modes in $z$ by eq. \casetwo\
and, exploiting the previous arguments,
the spurious poles at $w=\infty$ do not appear in eq. \propfin.
When $k=0$, instead, both tensors
$\phi(w)$ $f_0(z)$ are
linearly divergent at infinity. Fortunately, thanks to the term $1/(z-w)$,
$u_0(z,w)$ does not show up spurious poles in $z$ and $w$.
\smallskip
Finally, let us check that the singularity in $z=w$
of $K_\lambda(z,w)$ occurs only when $l=l'$
and it is a simple pole. The proof will be done without assuming
any restriction to the form of the Weierstrass polynomial \curve.
Let us start substituting in the expression
\ksum\ of $K_\lambda(z,w)$ the explicit form of $f_k$ and $\phi_k$ given
by the definitions \fkn\ and \phikn:
$$
K_{\lambda}(z,w)
=
{
[F_y(w,y(w))]^{\lambda -1}
 \over [F_y(z,y(z))]^{\lambda}
}
{1\over z-w}
dz^{\lambda}
dw^{1-\lambda}\times
$$
$$[y^{n-1}(z) + y^{n-2}(z)\left(
y(w)+P_{n-1}(w)\right) + y^{n-3}\left( y^2(w) + P_{n-1}(w)
y(w) + P_{n-2}(w)\right)$$
\eqn\super
{
+ ... + y^{n-1}(w) + y^{n-2}(w) P_{n-1}(w) + ... + P_{1}(w)]
}
where it is understood that $y(z)$ and $y(w)$ are in the branch $l$ and
$l'$ respectively.
It is now possible to write
$$ y^{n-1}(z) + y^{n-2}(z)\left(
y(w)+P_{n-1}(w)\right) + y^{n-3}\left( y^2(w) + P_{n-1}(w)
y(w) + P_{n-2}(w)\right)$$
\eqn\supera
{
+ ... + y^{n-1}(w) + y^{n-2}(w) P_{n-1}(w) + ... + P_{}(w)
={F(w,y(z))\over y(z) - y(w)}
}
Eq. \supera\ has been obtained multiplying and dividing the left hand side
by $y(z)-y(w)$ and then exploiting the formulas $F(w,y(w))=0$ and
$$a^n-b^n=(a-b)(b^{n-1}+ab^{n-2}+a^2b^{n-3}+\ldots+ a^{n-1})$$
With these simplifications it turns out that
\eqn\superb
{
K_{\lambda}^{(ll')} (z,w) = {[ F_y(w,y^{l'}(w))]^{\lambda -1}
 \over [F_y(z,y^l(z))]^{\lambda} } {1\over z-w}
{F(w,y^l(z))\over y^l(z) - y^{l'}(w)}dz^{\lambda}dw^{1-\lambda}
}
But this is the well known form of the Weierstrass
kernel on algebraic curves (see for instance ref. \ffstr) which has exactly
one pole in $z=w$ for $l=l'$ and spurious poles in $w=\infty$.
Since the latter spurious poles are irrelevant in the final expression of
the propagator as previously shown, we conclude that eq. \propfin\
yields the correct two point function of the $b-c$ systems with
$\lambda>1$ on the algebraic
curves of the kind \curve.
Eq. \superb\ shows that
the
Weierstrass polynomial is a superposition of the
multivalued modes $f_k(z)$ and $\phi_k(w)$ and it is a crucial result
for the operator formalism.
As a matter of fact, once
the Weierstrass polynomial and the zero modes are given,
it is possible to construct the correlation of the $b-c$ systems
automatically using the formulas given below. It is thus remarkable
that both the zero modes and the Weierstrass kernel exhibit the
splitting into $n$ sectors characterized by the tensors
$f_k$ and $\phi_k$, $k=0,\ldots,n-1$.
\smallskip
Starting from eq. \propfin, we are able to compute all the other
$n-$point functions applying the Wick theorem. The Wick theorem for the
$b-c$ systems has been rigorously studied in
\ref\raina{
A. K. Raina, {\it Helv. Phys. Acta} {\bf 63} (1990), 694.}
and it
is valid also in our case.
One can check it inductively. As previously shown,
the Wick theorem
based on the normal ordering \normord\ holds in the case of the two
point function \propfin. Now let us suppose
that the Wick theorem has been verified for the correlator
$$G_{N-1,M-1}(z_1,\ldots,z_{N-1};w_1,\ldots,w_{M-1})=\langle
0| b(z_1)\ldots  b(z_{N-1}) c(w_1)\ldots
 c(w_{M-1})|0\rangle$$
with $N-M=N_b$. Then, using eq. \normord\ we obtain
for $G_{N,M}(z_1,\ldots,z_N;w_1,\ldots,w_M)$:
$$\langle 0| b(z_N) c(w_M)
 b(z_1)\ldots  b(z_{N-1}) c(w_1)\ldots
 c(w_{M-1})|0\rangle=$$
\eqn\brak{\sum\limits_{i=1}^M(-1)^i K_\lambda(z_N,w_i)
G_{N-1,M-1}(z_1,\ldots,z_{N-1};w_1,\ldots,w_{i-1},w_{i+1},\ldots w_{M-1}
).}
All the other possible contractions vanish due to the fact that the Wick
theorem holds by hypothesis in the case of any product containing $N-1$ fields
$b$ and $M-1$ fields $c$.
As an upshot we obtain:
$$
<\prod\limits_{s=1}^M b^{(l_s)}(z_\rho)\prod\limits_{t=1}^N
 c^{(l'_t)}(w_t)>=$$
\eqn\bclambda{{\rm det}\left|\matrix{\Omega_1^{(l_1)}(z_1)&\ldots&
\Omega_{N_b}^{(l_1)}(z_1)&K_\lambda^{(l_1l'_1)}(z_1,w_1)&
\ldots&K_\lambda^{(l_1l_N')}(z_1,w_N)\cr
\vdots&\ddots&\vdots&\vdots&\ddots&\vdots\cr
\Omega_1^{(l_M)}(z_M)&\ldots&
\Omega_{N_b}^{(l_M)}(z_M)&K_\lambda^{(l_Ml'_1)}(z_M,w_1)&
\ldots&K_\lambda^{(l_Ml_N')}(z_M,w_N)\cr}\right |}
where $M-N=(2\lambda-1)(g-1)=N_b$.
The tensor $K_\lambda^{(ll')}(z,w)$ has spurious poles in the limit
$w\rightarrow\infty$. However one can show as before
that these poles do not contribute to the determinant \bclambda.
The important fact to be noted here is that both eqs. \propfin\ and
\bclambda\ were evaluated using a pure operator formalism on the complex
plane. This operator formalism, however, reproduces exactly the correlation
functions of the $b-c$ systems on general algebraic curves.
Exploiting the correspondence between algebraic curves and Riemann surfaces,
we can also say that eqs. \propfin\ and \bclambda\ represent the explicit
form of the formulas given
in refs. \vv.
\vskip 1cm
\newsec{THE OPERATOR FORMALISM ON A GENERAL QUARTIC OF GENUS THREE
AND ON THE $Z_n$ SYMMETRIC CURVES}
\vskip 1cm
In this section we treat two relatively simple cases of algebraic curves
for which the results can be compared with those obtained by applying
other methods \janstr, \ffstr, \knirev, \brzn, \ref\ferbos{F. Ferrari, {\it
Jour. Math. Phys.} {\bf 32} (1991), 2186.}.
The first example is the quartic $Q_3$ of genus three with
Weierstrass polynomial:
\eqn\quartic{y^3+3p(z)y-2q(z)=0}
where $p(z)$ and $q(z)$ are polynomials of degree three and four
respectively.
This curve represents all the possible Riemann surfaces of genus three
that are nonhyperellitpic.
$y(z)$ has nine finite branch points $a_1,\ldots,a_9$, which are
roots of the equation $p^3(z)+q^2(z)=0$.
Moreover, there is one branch point at infinity of multiplicity two.
By analysing this example
we demonstrate that the operator formalism applies also
to curves with branch points at infinity and only for technical reasons
we considered in the general discussion
the more restricted case in which
all the branch points are finite.
As a convention, we
suppose that the the $0-$th and $1-$st
sheets are joined at infinity, so that
$\infty_0=\infty_1=\infty_{01}$.
The relevant divisors are \ffstr:
\eqn\divdzqt{[dz]=\sum_{p=1}^9a_p-3\infty_{01}-2\infty_2}
\eqn\divyqt{[y]=\sum_{j=1}^4q_j-3\infty_{01}-\infty_2}
\eqn\divfyqt{[F_y]=\sum_{p=1}^9a_p-6\infty_{01}-3\infty_2}
where $F_y(z,y(z))=3(y(z)+p(z))$.
Now we notice that the Weierstrass polynomial \quartic\ can be
obtained from eq. \curve\ after setting:
$$P_4(z)=P_2(z)=0\qquad\qquad P_3(z)=1\qquad\qquad
P_1(z)=p(z)\qquad\qquad P_0(z)=-2q(z)$$
Since the generalized Laurent
expansions \bdz\ and \cdz\ are valid for any algebraic
curve, it is sufficient to perform the above substitution in eqs.
\bdz-\phikn, with $k=0,1,2$.
As a consequence,
the multivalued modes $f_k$ and $\phi_k$
become on $Q_3$:
\eqn\fkqt{
f_k(z)=
{
y^{2-k}(z)dz^\lambda\over[F_y(z,y(z))]^\lambda
}\qquad
\qquad\qquad k=0,1,2
}
\eqn\phiknqtzo{
\phi_k(w)=
{
y^k(w)dw^{1-\lambda}
\over [F_y(w,y(w))]^{1-\lambda}
}
\qquad\qquad\qquad k=0,1
}
\eqn\phiknqtt{
\phi_2(w)=
{
(y^2(w)+3p(w))dw^{1-\lambda}
\over[F_y(w,y(w))]^{1-\lambda}}
}
At this point we start to work out the operator formalism for $\lambda=2$.
In this way the notations are simpler and it is possible to better
illustrate
the way in which the operator formalism works, carrying out all the
calculations in details.
Apart from the evaluation of the relevant divisors,
the particular case under consideration is just a subcase of the general
operator formalism of Section 3 with $n=3$ and $\lambda=2$, so that
the space of the $b-c$ fields on $Q_3$
is splitted into three independent Hilbert spaces.
Therefore, we can exploit without modifications
eqs. \commrel-\vaccond\
in order to compute the correlation functions of the $b-c$ systems.
The only difference from section 3 is in the number
of zero modes, for which we have to use the divisors \divdzqt-\divfyqt.
After a straightforward calculation, we get the  explicit expression of
the six independent
quadratic differentials on $Q_3$:
$$\Omega_{0,0}=f_0(z)$$
$$\Omega_{1,j}(z)dz^2=z^{j-1}f_1(z)\qquad\qquad\qquad j=1,2$$
$$\Omega_{2,j}dz^2=z^{j-1}f_2(z)\qquad\qquad\qquad j=1,2,3$$
As a consequence the number of zero modes in the different $k-$sectors becomes:
\eqn\zmqtn{N_{b_0}=1\qquad\qquad N_{b_1}=2\qquad\qquad N_{b_2}=3}
\smallskip
Now we are ready to prove eq. \proptwo.
To this purpose,
we start from the simplest nonvanishing correlator \proptwo, which contains
in this case six zero modes:
\eqn\start{\langle0| b(z_1)\ldots  b(z_6)|0\rangle\equiv
\langle0|\prod\limits_{i=1}^6\left(\sum\limits_{k=0}^2b_k(z_i)\right)
|0\rangle}
The product in the right hand side of \start\ can be expanded
in all possible monomials of the fields $b_k$
using Lemma 1 of Appendix B. Due to the condition \vaccond, only the
monomials containing three fields $b_2$, two fields $b_1$ and one field $b_0$
survive after normal ordering. It is straightforward to check that these
monomials are given by the following formula:
$$\langle0| b(z_1)\ldots
b(z_6)|0\rangle=$$
\eqn\intone{\sum\limits_\sigma {\rm sign}(\sigma)\enskip_2\langle0|
\prod\limits_{l_2=1}^{3}b_2(z_{\sigma(l_2)})|0\rangle_2
\enskip_1\langle0|\prod\limits_{l_1=4}^{5}
b_1(z_{\sigma(l_1)})|0\rangle_1\enskip_0\langle0|
b_0(z_{\sigma(6)})|0\rangle_0}
where the sum
over $\sigma$ runs over all permutation of the integers
$1,\ldots,6$, ordered in such a way that:
$$\sigma(1)<\sigma(2)<\sigma(3)\qquad\qquad\sigma(4)<\sigma(5)$$
The vacuum expectation values remaining in eq. \intone\
can be directly computed
from the definitions \bdz\ and \vaccond\ or using eq. \propone.
In either cases the result is:
$$
\langle0| b(z_1)\ldots
b(z_6)|0\rangle=\sum\limits_\sigma {\rm sign}(\sigma){\rm det}\left|\Omega_{
2,j_2}(z_{\sigma(l_2)})\right|{\rm det}\left|\Omega_{
1,j_1}(z_{\sigma(l_1)})\right|\Omega_{0,1}(z_{\sigma(6)})$$
with $j_2,l_2=1,2,3$ and $j_1,l_1=4,5$.
It is now easy to recognize that the right hand side of the above equation
is the determinant of the six holomorphic
quadratic differentials,
taken at the points $z_1,\ldots,z_6$ and expanded as shown in Lemma 2
of Appendix B.
This concludes the proof of eq. \proptwo\ on the curve $Q_3$:
\eqn\proptqt{\langle0| b(z_1)\ldots
b(z_6)|0\rangle={\rm det}\left|\Omega_I(z_J)\right|}
Finally, we compute the two point function.
The creation and annihilation operators are defined as in eqs. \ban-\ccrea\
after the substitutions $\lambda=2$ and $n=3$. Moreover,
the number of zero modes in the
different $k-$sectors is given by eq. \zmqtn.
As a consequence, the normal ordering \normord\ becomes:
\eqn\normoqt{
 b(z)
c(w)=: b(z) c_k(w):+K_2(z,w)dz^2dw^{-1}
}
{}From the definition \ksum\ of $K_2(z,w)$ we have
$$K_2(z,w)dz^2dw^{-1}=
{dz^2dw^{-1}\over z-w}
\left[{F_y(w,y(w))\over F_y(z,y(z))}\right]\left(
{y^2(z)+y(z)y(w)+y^2(w)+3p(w)\over F_y(z,y(z))}\right)$$
The final form of $K_2(z,w)dz^2dw^{-1}$ is obtained
multiplying and dividing the right hand side of
the above equation by $y(z)-y(w)$:
\eqn\wkqt{
K_2(z,w)dz^2dw^{-1}=
{
dz^2dw^{-1}\over z-w
}
\left[{
F_y(w,y(w))\over F_y(z,y(z))
}\right]
{
F(w,y(z))\over (y(z)-y(w))F_y(z,y(z))
}
}
where $F(z,y(z))$ denotes the Weierstrass polynomial \quartic.
\smallskip\noindent
The proof that $K_2^{(ll')}(z,w)dz^2dw^{-1}$
is a good Weierstrass kernel with only one simple pole
in $z=w$ and $l=l'$ was already given in ref. \ffstr\ and we do not report
it here.
Since there are no spurious poles, we can compute the two point
function of the $b-c$ system with $\lambda=2$ using the Wick theorem
as shown in eq. \propint:
$${\langle 0| b^{(l)}(z)
c^{(l')}(w)\prod\limits_{I=1}^{6} b^{(l_I)}(z_I)
|0\rangle\over
\langle 0|\prod\limits_{I=1}^{6} b^{(l_I)}(z_I)|0\rangle}=$$
\eqn\propfinqt{{
{\rm det}\left|\matrix{\Omega_1^{(l)}(z)&\ldots&
\Omega_{6}^{(l)}(z)&K_2^{(ll')}(z,w)\cr
\Omega_1^{(l_1)}(z_1)&\ldots&
\Omega_{3}^{(l_1)}(z_1)&K_2^{(l_1l')}(z_1,w)\cr
\vdots&\ddots&\vdots&\vdots\cr
\Omega_1^{(l_{3})}(z_{3})&\ldots&
\Omega_{3}^{(l_{3})}(z_{3})&K_2^{(l_{3}l')}(z_{3},w)
\cr}\right |\over
{\rm det}\left|\Omega_I(z_J)\right|}}
with $I,J=1,\ldots 6$.
This is exactly the explicit propagator of the ghost of string theory
derived in ref. \ffstr. In order to derive also the higher order correlation
functions it is just sufficient to
exploit the Wick theorem as explained in Section 3.\smallskip
Now we pass to the $Z_n$ curves
of the kind:
\eqn\zncurve{y^n=\prod_{i=1}^{nm}(z-a_i)} where
$n$ and $m$ are integers.
Strictly speaking,
this class of curves is
degenerate in the sense of eq. \degeneracy. It is thus interesting
to verify through this example that the operator
formalism applies also to more general curves than
those treated in section 2.\smallskip
The $Z_n$ curves can be obtained from eq. \curve\
after the following substitutions:
$P_n(z)=1$, $P_j(z)=0$, $j=1,\ldots,n-1$ and $P_0(z)=
\prod_{i=1}^{nm}(z-a_i)$.
The points $a_i\in{\rm\bf C}$ are the branch points of the curve, so that
$n_{bp}=nm$.
The genus of the $Z_n$ curves \zncurve\ is given by:
\eqn\genus{g=1-n+{nm(n-1)\over 2}.}
and the relevant divisors are:
$$[dz]=(n-1)\sum_{p=1}^{nm}a_p-2\sum_{j=0}^{n-1}\infty_j$$
$$[y]=\sum_{p=1}^{nm}a_p-m\sum_{j=0}^{n-1}\infty_j$$
$$[F_y]=(n-1)\sum_{p=1}^{nm}a_p-(n-1)m\sum_{j=0}^{n-1}\infty_j$$
For simplicity, we consider here
only Riemann surfaces of genus $g\ge 2$, i.e. neither
the torus nor the sphere, for which an
operator formalism is already known.
The generalized Laurent expansions \bdz\ and \cdz\ become in this
case:
\eqn\bdz{b(z)dz^\lambda=\sum\limits_{k=0}^{n-1}\sum\limits_{i=-\infty}
^\infty b_{k,i}z^{-i-\lambda}f_k(z)dz^\lambda}
\eqn\cdz{c(z)dz^{1-\lambda}=
\sum\limits_{k=0}^{n-1}\sum\limits_{i=-\infty}^\infty
 c_{k,i}z^{-i+\lambda-1}\phi_k(z)dz^{1-\lambda}}
where
\eqn\znfkn{f_k(z)dz^\lambda ={dz^\lambda\over[y(z)]^{-k+\lambda(n-1)}}
\qquad\qquad\qquad k=0,\ldots,n-1}
\eqn\znphikn{\phi_k(z)dz^{1-\lambda}=
{dz^{1-\lambda}\over[y(z)]^{k-\lambda(n-1)}}\qquad
\qquad\qquad k=0,\ldots,n-1.}
The last ingredient of the operator formalism is provided by the zero modes,
which are of the form:
\eqn\zmzn{\Omega_{k,j}dz^\lambda={z^{j-1}dz^\lambda
\over[y(z)]^{-k+\lambda(n-1)}}\qquad\qquad\qquad j=1,\ldots,N_{b_k}}
where $N_{b_k}=-2\lambda+1+\lambda(n-1)m-km$.
It is easy to check that
$\sum\limits_{k=0}^{n-1}N_{b_k}=(2\lambda-1)(g-1)$, giving exactly the
number of the zero modes predicted by the
Riemann-Roch theorem. A treatment of the $b-c$ systems on the $Z_n$ curves
can be now performed using the techniques explained in Section 3.
The creation and
annihilation operators are defined by the following relations:
\eqn\ban{b^-_{k,i}|0\rangle_k\equiv b_{k,i}|0\rangle_k=0
\qquad\qquad\qquad\left\{\eqalign{k=&0,\ldots,n-1\cr i
\ge& 1-\lambda\cr}\right.}
\eqn\can{c^-_{k,i}|0\rangle_k\equiv c_{k,i}|0\rangle_k=0
\qquad\qquad\qquad\left\{\eqalign{k=&0,\ldots,n-1\cr i\ge&
\lambda\cr}\right.}
\eqn\bcrea{{}_k\langle0| b^+_{k,i}\equiv {}_k\langle 0|b_{k,i}=0
\qquad\qquad\qquad\left\{\eqalign{k=&0,\ldots,n-1\cr i
\le& -\lambda-N_{b_k}\cr}\right.}
\eqn\ccrea{{}_k\langle 0|c^+_{k,i}\equiv {}_k\langle 0|c_{k,i}=0
\qquad\qquad\qquad\left\{\eqalign{k=&0,\ldots,n-1\cr i\le&
\lambda-1\cr}\right.}
Finally, the normal ordering of the product of two fields
$b(z)c(w)$ is given by:
\eqn\nozn{b_k(z)c_k(w)=:b_k(z)c_k(w):+{1\over z-w}f_k(z)\phi_k(w)}
\eqn\noznp{c_k(z)b_k(w)=:c_k(z)b_k(w):+{1\over z-w}f_k(w)\phi_k(z)
.}
so that the Weierstrass kernel takes the form:
\eqn\wkzn{K_\lambda(z,w)dz^\lambda dw^{1-\lambda}=
{dz^\lambda dw^{1-\lambda}\over z-w}\sum_{k=0}^{n-1}\left[{y(w)\over y(z)}
\right]^{\lambda(n-1)-k}}
Moreover, in order to take into account also the zero modes, we
impose again the  condition \vaccond\ on the vacua $|0\rangle_k$:
$${}_k\langle 0|0\rangle_k=0\qquad{\rm if}\ N_{b_k}\ne 0;
\qquad\qquad
{}_k\langle 0|\prod\limits_{i=1}^{N_{b_k}}b_{k,i}|0\rangle_k=1.$$
With the above definitions, we can easily prove
eqs. \propone\ and \proptwo\ using the formulas of Appendix B.
Exploiting the normal ordering \nozn\ and \noznp\ we obtain the final formula
of the propagator in a purely operatorial way:
$${\langle 0|b^{(l)}(z)c^{(l')}(w)\prod\limits_{I=1}^{N_b}b^{(l_I)}(z_I)
|0\rangle\over
\langle 0|\prod\limits_{I=1}^{N_b}b^{(l_I)}(z_I)|0\rangle}=$$
\eqn\propfin{{
{\rm det}\left|\matrix{\Omega_1^{(l)}(z)&\ldots&
\Omega_{N_b}^{(l)}(z)&K_\lambda^{(ll')}(z,w)\cr
\Omega_1^{(l_1)}(z_1)&\ldots&
\Omega_{N_b}^{(l_1)}(z_1)&K_\lambda^{(l_1l')}(z_1,w)\cr
\vdots&\ddots&\vdots&\vdots\cr
\Omega_1^{(l_{N_b})}(z_{N_b})&\ldots&
\Omega_{N_b}^{(l_{N_b})}(z_{N_b})&K_\lambda^{(l_{N_b}l')}(z_{N_b},w)
\cr}\right |\over
{\rm det}\left|\Omega_I(z_J)\right|}.}
where the Weierstrass kernel $K_\lambda(z,w)$ has been defined in eq. \wkzn.
Also in the case of the $Z_n$ curves, one can easily show that the spurious
poles of $K_\lambda(z,w)$ at $w=\infty$ are harmless. Finally the higher
order correlation functions can be obtained as in the general case from the
Wick theorem. Formally, they look like those in eq. \bclambda\ apart from the
fact that the Weierstrass kernel is that of eq. \wkzn\ and the zero modes are
those of eq. \zmzn.
The results obtained with the operator formalism in the $Z_n$ case can be
compared with the analogous results obtained in a purely geometrical way
in ref.
\ferbos.
\vskip 1cm
\newsec{CONCLUSIONS}
\vskip 1cm
We have seen in this paper that the correlation functions of the $b-c$ systems
with integer $\lambda>1$ can be computed on any nondegenerate
algebraic curve
using a pure operator formalism.
However, we would like to stress that our
operator formalism applies to arbitrary algebraic
curves. This claim has been verified for instance
in the case of the $Z_n$ algebraic
curves \zncurve\ which
are degenerate with respect to to the definition
\degeneracy\ given in section 2. Moreover, in section 4 we have also
considered the example of the quartic $Q_3$ with a branch point at infinity.
The essential of our construction is based on the fact that
the relevant modes propagating
in the amplitudes are multivalued fields defined on the complex plane.
Crucial points of our approach are the generalized Laurent expansion
explained in Section 2 and the splitting of the space of the $b-c$ fields
into $n$ independent Hilbert spaces.
The splitting directly follows from the fact that both the zero modes
and the Weierstrass kernel, which are the main ingredients
in order to construct the correlation functions (see eq. \bclambda),
are superpositions
of the multivalued modes $f_k(z)$ and $\phi_k(w)$ given in
eqs. \fkn\ and \phikn. For the zero modes, this is a straightforward
consequence of
Proposition 1. For the Weierstrass kernel, instead, this has been proved
in
eq. \superb.
In both cases, no assumption on the form of the Weierstrass
polynomial was made.
As an upshot, eq. \bclambda\ provides the explicit form of
the correlation functions of the $b-c$ systems on arbitrary
closed and orientable Riemann surfaces.
Once the Weierstrass polynomial is given,
the Weierstrass kernel $K_\lambda(z,w)$ can be computed from eq. \superb\
in terms of the coordinates $z$ and $y$.
The derivation of the zero modes, instead, is more complicated,
since there are no universal procedures which yield the
zero modes on any algebraic curve without previously knowing the
divisors of the basic building blocks $dz$, $y$ and $F_y$.
This is just a formal problem because, once the form of
the Weierstrass polynomial is fixed, we are able to obtain
the above divisors exploiting the techniques of ref. \ffstr.
However,
due to the fact that
it is not possible to cover in one paper all the existing families of curves,
the explicit calculation of the zero modes and the proof
that there are not spurious singularities in the amplitudes has been performed
after imposing some assumptions on
the Weierstrass polynomial.
The chosen families of curves, however, are very general and do not
exhibit any discrete group of symmetry.
\smallskip
Beyond the already discussed results in string theory, we would like
also to outline some interesting consequences
following from our analysis in the study
of conformal field theories and of the Riemann monodromy
problem.
To begin, we notice that the
Laurent bases described in Section 2 represent
a great advantage with respect to the usual Poiseaux series, which allow
only local expansions of the multivalued analytic tensors.
For instance, representing the variable $z$ in polar coordinates,
the bases \basisone\ and \basistwo\ can be regarded
as the global generalization of the usual Fourier series.
The possibility of solving differential equations on any algebraic curve,
also with boundary,
by means of these
series is currently under
investigation.
Moreover, starting from eq. \fkn\ and putting $\lambda=0$, we are
able to reproduce
every multivalued function defined by a general Weierstrass polynomial.
In this sense the Riemann monodromy problem \ref\rimopr{J. Plemelj,
Problems in the sense of Riemann and Klein,
Interscience, New York, 1964; G. D. Birkhoff, Collected Mathematical Papers I,
Dover, New York, 1968; I. A. Lappo-Danilewski, M\'emoires sur la th\'eorie des
syst\'emes des \'equations differ\'entielles lin\'eaires, Chelsea,
1953; M. Muskhelishvili, Singular integral equations, Groningen, 1953.},
intended here as
the problem of finding all the independent
functions multivalued according to the monodromy
data given by the branch points and the local monodromy group of an
arbitrary algebraic curve, has been implicitly solved.
For example, let us consider one of the simplest examples,
the quartic $Q_3$ of genus three given in eq. \quartic.
In this case it is known that the Riemann monodromy problem
admits three independent solutions whose zeros and poles must be
concentrated at the branch points and at $z=\infty$ \rimopr.
Two rationally independent functions fulfilling these constraints are
$f_0(z)=1$, the zero mode, and $f_2(z)=F_y(z,y(z))$ while,
by Proposition 1, the third function should have the form
given in eq. (A.6) for $\lambda=0$:
$$f_1(z)=g_0(z)+g_1(z)y(z)+g_2(z)y^2(z)$$
After a few calculations, one finds:
$$f_1(z)=p(z)^2+2q(z)y-p(z)y^2$$
As a matter of fact, $f_1(z)$ is proportional to the square
of $f_2(z)$ which, as already remarked, has all its
zeros at the branch points and the poles at infinity.
Thus, once the form of the Weierstrass polynomial is known, the problem
of constructing a complete set of independent functions with given zeros and
poles at the branch points can be easily solved.
The difficulties appear instead
in reducing the differential equations
satisfied by these functions to the standard form of the Riemann monodromy
problem\foot{
Conversely, in the usual methods of solving the
Riemann monodromy problem, \rimopr,
\ref\sjm{
M. Sato, T. Miwa and M. Jimbo. Holonomic quantum fields
(Kyoto U.P. Kyoto), part I; 14 (1978) p. 223; II: 15 (1979) p. 201;
III: 15 (1979) p. 577; IV: 15 (1979) p. 871; V; 16 (1980) p.
531.} the equations are in the standard form
from the beginning, but the solutions are in the
form of
infinite series. When no exact convergence is guaranteed,
the independent solutions are valid only locally and should be understood
in terms of analytic continuation.}.
\smallskip
The reason for which the differential equations satisfied by a rational
function $f(z,y)$ of $z$ and $y$ become complicated is that $f(z,y)$
is defined on a Riemann
surface. Therefore, the two dimensional Poincar\'e group
is explicitly broken.
To explain this, we consider the
equations of the Riemann monodromy problem
as flux equations with complex time $z$:
\eqn\rmop{{d\Psi(z)\over dz}=v(z)\Psi(z)}
where $\Psi(z)$ is a vector of multivalued functions and
$v(z)\equiv v_z(z)$ is a matrix of (possibly multivalued) vectors.
We remember that the form of $v(z)$ is
constrained by the requirement that all its singularities
should be simple poles at the branch points and
in $z=\infty$.
It is now easy to see that on the complex plane, as well as on the
$Z_n$ symmetric curves, the only way of fulfilling this condition
is given by:
\eqn\standard{v_{ij}(z)dz={A_{ij}dz\over z-a_j}}
where the $a_j$ represent the branch points and the $A_{ij}$ are constants.
Thus the vector $v(z)$ is invariant under the global translations of the
coordinates $z$ and $a_j$.
On a general algebraic curve, however, eq. \standard\ is not the only
possibility.
For example, using the divisors \divdz-\divdf\
we can immediately check that the following vector has only simple poles
at the branch points:
\eqn\vnnn{v_z(z)={dz\over [F_y(z,y(z))]^2}}
Vectors of the kind \vnnn\ break explicitly
the invariance under translations exhibited by eq. \rmop\ and
lead in principle to nonlinear monodromy
equations which seems to have not much in common with the
familiar Fuchsian equations on the complex plane.
Of course, these nonlinear equations must be reduced to the form
\standard, but this is not easy to achieve, since on a
general algebraic curve there are no explicit
functional relations between the branch points and the parameters appearing
in the Weierstrass polynomial.
Until now, the only class of algebraic curves with nonabelian
group of symmetry for which the solutions of the Riemann monodromy
problem have been explicitly constructed in terms
of $z$ and $y(z)$ are the $D_n$ curves of ref.
\ref\ffbgs{F. Ferrari, {\it Int. Jour. Mod. Phys.} {\bf A9} (1994), 313.}.
The results obtained in \ffbgs\ confirm that the derivatives with respect
to $z$
of these solutions contain vectors of
the form \vnnn. After some manipulations, however, the Riemann monodromy
problem can be reduced
to a set of equations given by Plemelj - see the
first reference of
\rimopr\
- which is equivalent to the usual Fuchsian
formulation \standard.\smallskip
Concluding, the techniques explained in this work
open the possibility of
performing explicit calculations on any Riemann surface.
In this way, the correlation functions of the $b-c$ systems
with $\lambda>1$ have been computed for the first time without bosonization
and the Hilbert space of the $b-c$ fields has been simply reproduced
in terms of multivalued modes on the complex plane.
It would be interesting to study with the same methods
the Hilbert space of massless scalar fields. This case is however
complicated by the fact that the theory is not completely conformal
and there is a mixing between $z$ and its complex conjugate
variable $\bar z$ in the amplitudes.
The applications in the Riemann monodromy problem
discussed before
are also relevant in the case of the
Knizhnik-Zamolodchikov equations, which are of the same form \rmop.
Also in the Knizhnik-Zamolodchikov
equations,
vectors of the kind \vnnn\ appear
due to the explicit breaking of the two dimensional Poincar\'e
group on a Riemann surface. The way in which
this affects the particle statistics  inside the amplitudes
has been studied on the $D_n$ curves in \ffbgs, but the
analysis on more general curves is still missing.
Our results provide powerful tools in order to simplify these equations
and we hope that in this way it will be possible
to obtain
new insights on the interplay between algebraic curves,
integrable models and Riemann monodromy problem.
\smallskip
\vskip 1cm
\appendix{A}{}
\vskip 1cm
In this Appendix we prove Proposition 1 of Section 2.
To this purpose, let us consider a general $\lambda-$differential
$\omega dz^\lambda$ defined on an arbitrary algebraic curve described
by the two coordinates $z$ and $y(z)$, where $y(z)$ satisfies eq. \curve.
As it is well known, $\omega dz^\lambda$ should be of the form:
\eqn\omedef{\omega dz^\lambda ={Q(z,y(z))\over R(z,y(z))}dz^\lambda}
where $Q$ and $R$ are polynomials in $z$ and $y(z)$.
Moreover, using eq. \curve\
it is always possible to make the degree of $P$ and $Q$ in the
variable
$y(z)$ equal to $n-1$ or lower.
In fact, every polynomial containing powers $y^m$ with $m\ge n$
can be transformed in a polynomial of lower degree
in the obvious way:
\eqn\reduction{y^m=-\left[P_{n-1}(z)y^{m-1}+P_{n-2}(z)y^{m-2}+\ldots+P_0(z)
y^{m-n}\right]}
Now we show that any $\lambda-$differential $\omega dz^\lambda$
admits the following expansion:
\eqn\expanint{\omega dz^\lambda=\sum\limits_{k=0}^{n-1}g_k(z)y^k(z)dz^\lambda}
where the $g_k(z)$ are rational functions of $z$.
To this purpose, it is sufficient to rewrite the denominator $1/R(z,y(z))$
appearing in the definition of $\omega dz^\lambda$ as a polynomial in $z$ and
$y$.
$R(z,y)$ is of the form:
\eqn\rdef{R(z,y)=\sum\limits_{i=1}^n R_{n-i}(z)y^{n-1}}
where the $R_{n-i}(z)$ are polynomials in the variable $z$
only.
Now we set $\tilde y\equiv R(z,y(z))$. $\tilde y$ is a multivalued function
with the monodromy properties of $y$. By eq. \reduction, the
powers $\tilde y,\tilde y^2,\ldots,\tilde y^n$ remain polynomials
in $y$ and $z$ at most of order $n-1$ in $y$.
As a consequence, there must exist $n+1$ polynomials $\tilde P_i(z)$,
$i=0,\ldots,n$, such that the following Weierstrass polynomial is
satisfied:
\eqn\curvetilde{\tilde P_n(z)\tilde y^n+\tilde P_{n-1}(z)\tilde
y^{n-1}+\ldots+\tilde P_1(z)\tilde y+\tilde P_0(z)=0}
We notice that in some very peculiar cases, the polynomial
$R(z,y)$ may not faithfully represent the local monodromy group $G$
of $y$. In this case, the local monodromy group of $\tilde y$
is a subgroup of $G$ and the degree of the polynomial \curvetilde\ in
$\tilde y$ is lower than $n$.
We exclude also the possibility that $\tilde P_0=0$, because it is easy
to see that this would imply that the Weierstrass polynomial
\curvetilde\ is reducible and one of its solutions is $\tilde y=0$.
But this cannot be true because, from the beginning, we suppose
that the denominator of eq. \omedef\ does not identically vanish.
Remembering that $R(z,y(z))=\tilde y$ by definition and using
eq. \curvetilde, it is now easy to see that:
$$\omega dz^\lambda=-{Q(z,y(z))\over \tilde P_0(z)}\left[\tilde
P_n(z)R^{n-1}+\tilde P_{n-1}(z)R^{n-2}+\ldots+\tilde
P_1(z)\right]dz^\lambda$$
where $R$ has been defined in eq. \rdef. It is then clear that
the right hand side
of the above equation is a finite linear combination containing
positive powers of $y$. The coefficients of the combination are
rational functions of $z$. Therefore it is easy to bring $\omega
dz^\lambda$ in the desired form \expanint\
using again eq. \reduction\ iteratively.
\smallskip
Eq. \expanint\ means that an arbitrary meromorphic
$\lambda-$differential can be expanded in the generalized
Laurent basis whose elements are given by:
\eqn\genbasis{
G_{i,k}(z,y(z))
dz^\lambda=z^i y^k(z)dz^\lambda\qquad\qquad\qquad\cases{i=0,1,\ldots\cr
k=0,\ldots,n-1\cr}}
To complete the proof of Proposition 1, we just need to show that the
basis \basisone\ is equivalent to that of eq. \genbasis.
Indeed, the elements
$B_{i,k}(z,y(z))
dz^\lambda$ of eq. \basisone\ are $\lambda-$differentials of the form \omedef\
and can be rewritten as linear combinations of the
$G_{i,k}(z,y(z)) dz^\lambda$
as in eq. \expanint.
Conversely, given an element
$G_{i,k}(z,y(z))dz^\lambda$ it is always possible to express it as a
linear combination of the
$B_{i,k}(z,y(z))dz^\lambda$. As a matter of fact, the equation
\eqn\equivalency{G_{i,k}(z,y(z)) dz^\lambda=\sum\limits_{i'\in{\rm\bf Z}}
\sum\limits_{k'=0}^{n-1}a_{ii'}b_{kk'}B_{i',k'}(z,y(z))
dz^\lambda}
where $a_{ii'}$ and $b_{kk'}$ are constants, amounts to the following
identity:
$$[F_y(z,y)]^\lambda z^i y^k=
\sum\limits_{i'\in{\rm\bf Z}}\sum\limits_{k'=0}^{n-1}a_{ii'}b_{kk'}
z^{i'} y^{k'}$$
But $[F_y(z,y)]^\lambda$ is again a polynomial in $z$ and $y$ which
can be reduced by means of eq. \reduction, so that there exists a finite
number of constants $a_{ii'}$ and $b_{kk'}$
satisfying eq. \equivalency.
This concludes the proof of Proposition 1. An analogous result can be
found for the basis of the $c$ fields \basistwo.
\smallskip
\vskip 1cm
\appendix {B}{}
\vskip 1cm
The proof of the eq. \proptwo\ becomes quite straigthforward once two
basic lemmas are established.\medskip\noindent
{\bf Lemma 1}\smallskip
\eqn\super{\prod\limits_{i=1}^N (\sum\limits_{j=1}^Ma_{i,j})=
\sum\limits_{r_1+\ldots +r_M=N}
\sum\limits_{\sigma}{\rm sgn}(\sigma )\prod\limits_{k=1}^M\prod\limits_
{l_k=\alpha (k)}^{\beta (k)}a_{\sigma (l_k),k}}
where $a_{k,j}$ are anticommuting Grassmann variables,
\eqn\supexpl{r_j\geq 0;\quad r_0=0;\quad \alpha (k)=1+\sum\limits_
{m=1}^{k-1}r_m;\quad \beta (k)=\alpha (k)+r_k-1.}
The symbol $\sum\limits_{\sigma}$ in the \super\ denotes the sum over
all
the
permutations of numbers 1,...,N such that
\eqn\permutrestr{\sigma(\alpha (k))<\ldots <\sigma(\beta (k))\qquad
k=1,\ldots ,M.}
The simplest way to proof \super\ is to perform an induction in M.

\medskip\noindent
{\bf Lemma 2}\smallskip
Consider a $N\times N$ matrix $A$ with commuting
elements $a_{i,k}$.
$i$ is the row index, while $k$ counts the columns of $A$.
We note that, contrary to the notation exploited in the previous Lemma,
the elements $a_{i,k}$'s
of the matrix $A$ represent now non-Grassmann variables.
\smallskip
Suppose we
have a partition of $N$ into $M$ integers: $N=r_1+r_2+\ldots+r_M$ where
$r_j\geq 0$. For each permutation $\sigma$ satisfying the condition
\permutrestr\ we define matrices:
\eqn\kasigma{A_{\sigma}^{(k)}=\left(\matrix{
a_{\sigma (\alpha (k)),\alpha(k)}&\ldots&a_{\sigma(\alpha (k)),\beta(k)}\cr
\vdots&\ddots&\vdots\cr
a_{\sigma (\beta (k)),\alpha(k)}&\ldots&a_{\sigma (\beta (k)),\beta(k)}
\cr}\right)}
where $\alpha (k)$ and $\beta (k)$ are defined in \supexpl.
Then
\eqn\determinant{{\rm det}A=\sum\limits_\sigma {\rm sign}(\sigma)
\prod_{j=1}^M {\rm det}|A^{(j)}_\sigma| }
where $\sigma$ is an arbitrary permutation satisfying
\permutrestr .
This is a generalization of the usual way of computing determinants
(see for example \ref\nl{F. Nei\ss\
and H. Liermann, Determinanten und Matrizen,
8$^{\rm th}$ ediction, Springer Verlag, (in german).}).
The best known subcase of eq. \determinant\
is that for which $r_1=1$, $r_2=N-1$ and  the remaining $r_j$ vanish.

With the help of Lemmas 1 and 2 we can  check eq. \proptwo. Our aim
is to calculate the correlator $\langle 0|\prod\limits_{I=1}^{N_b}
b(z_I)|0\rangle$.
The strategy is to decompose the fields $b(z_I)$ according to
\bdzcdz.
We apply now the Lemma 1 and
obtain a sum over all possible partitions of the number $N_b$. According
to properties of the vacuum expressed in \vaccond\ only one partition
gives a nonzero contribution: the number of $b_k$ fields has to be
equal $N_{b_k}$. Once we establish this fact we arrive
immediately at the product of $n$ correlators of the type given in eq.
\propone. With the help of Lemma 2 we immediately complete the proof of
\proptwo.

\listrefs
\end